\documentclass[osajnl,twocolumn,showpacs,superscriptaddress,10pt]{revtex4-1} 
\usepackage{amsmath,amssymb,graphicx}

\usepackage{subfigure}
\usepackage{subeqnarray}
\usepackage{cases}
\usepackage{multirow}
\usepackage{caption2}
\usepackage{threeparttable}
\begin{document}

\title{Large phase shift of (1+1)-dimensional nonlocal spatial solitons in lead glass}

\author{Qian Shou}

\author{Miao Wu}
\author{Qi Guo}\email{Corresponding author: guoq@scnu.edu.cn}
\affiliation{Laboratory of Nanophotonic Functional Materials and Devices, South
China Normal University, Guangzhou 510006}

\begin{abstract}The large phase shift of strongly nonlocal spatial optical
soliton(SNSOS) in the (1+1)-dimensional [(1+1)D] lead glass is
investigated using the perturbation method. The fundamental soliton
solution of the nonlocal nonlinear Sch\"{o}dinger equation(NNLSE)
under the second approximation in strongly nonlocal case is
obtained. It is found that the phase shift rate along the
propagation direction of such soliton is proportional to the degree
of nonlocality, which indicates that one can realize
$\pi$-phase-shift within one Rayleigh distance in (1+1)D lead glass.
A full comprehension of the nonlocality-enhancement to the phase
shift rate of SNSOS is reached via quantitative comparisons of phase
shift rates in different nonlocal systems.
\end{abstract}


\maketitle 

\section{Introduction}
Nonlocal spatial solitons have been the subject of intensive
experimental and theoretical work \cite{Assanto-PRL-2004,
Rotschild-PRL-2005,Bang-PRE-2002,Peccianti-OL-2002} since the
pioneering work done by Snyder and Mitchell \cite{SM-Science-97}.
The most prominent innovation in their work is that they transforms
the complex nonlocal nonlinear Sch\"{o}dinger equation (NNLSE) into
a simple case of linear propagation of light in a quadratic
self-induced index well \cite{SM-Science-97}. Nonlocal nonlinearity
is typically the result of certain transport processes, such as the
charge drift in photorefractive crystals \cite{Galvo-EL-2002} and
the heat transfer in thermal nonlinear media
\cite{Rotschild-PRL-2005} or, long-range interaction, such as the
molecular reorientations in liquid crystals
\cite{Peccianti-PR-2012}. Due to the nature of the nonlocality,
solitons in nonlocal nonlinear media exhibit several distinct
properties that are not possible in local settings. This includes,
on one hand, resulting from the spatial `averaging' character of the
nonlocality, the arrest of catastrophic collapse
\cite{Bang-PRE-2002}, the ability to support the formation of
complex optical spatial solitons, such as higher-order solitons
\cite{Rotschild-OL-2006,Ye-PRA-2008} and vortex solitons
\cite{Rotschild-PRL-2005, Shen-JO-2012}. On the other hand,
out-of-phase solitons attraction
\cite{Rasmussen-PRE-2005,Hu-APL-2006}, long-range interactions
between solitons \cite{Rotschild-NP-2006} as well as the solitons
and the boundaries \cite{Alfassi-OL-2007,Shou-OL-2009,
Buccoliero-JOA-2009} in strongly nonlocal media have also been
carried out or predicted due to the fact that the interactions are
mediated by the light-induced refractive index which is `enlarged'
by the nonlocal response.

Except for the `averaging' and the `enlarging' features of the
nonlocality, there exists an `enhancing' effect of the nonlocality
on the phase shift of the SNSOS. Although very large in fact, the
phase shift of SNSOS is considered a trivial term, for a long time,
and is neglected by the Snyder-Mitchell (SM) model
\cite{SM-Science-97}. The first work focused on the phase shift of
SNSOS was done by Guo $et$ $al$. \cite{Guo-PRE-2004,Xie-oqe-2004}.
They predicted a large phase shift rate of SNSOS, which is
$\alpha^{2}$ times ($\alpha$ is the degree of nonlocality defined as
$\alpha =w_{m}/\mu$ where $w_{m}$ is the characteristic length of
the response function and $\mu$ is the beam width), explicitly 100
times for the lower limit of the strongly nonlocality, larger than
that of the local counterpart. Guo's conclusion results from a
strongly nonlocal (SN) model in which the large phase shift is
included having a dominating term proportional to the soliton
critical power.

SN model can rigorously transform to SM model with a function
transformation involving large phase shift term \cite{Guo-2013}. Both of them are derived from a
phenomenological and regular (or at least twice-differentialble at
$\textbf{r}=0$) response function $R(\textbf{r})$. In the nematic liquid crystal (NLC) and lead glass(LG),
the two media found so far in which SNSOSs can form, the response
functions are singular at every source point (irregular) and
therefore one can not obtain accurate solution of NNLSE based on SN
model and SM model even in the strongly nonlocal case\cite{Guo-2013}. Ouyang $et$ $al$. took the higher order (the
forth and the sixth) terms of the light induced refractive index as
the perturbation to the quadratic index well and obtained
considerably accurate analytical soliton solutions in (1+1)D
\cite{Ouyang-PRE-2006} and (1+2)D \cite{Ouyang-PRA-2007} NLC. The perturbation solution are different from
Gaussian-type solution given by SN model\cite{Guo-2013}, but still
indicated nonlocality-enhanced large phase shifts of SNSOSs. The
first theoretical and experimental study focused on the SNSOS phase
shift was carried out in (1+2)D cylindrical LG by Shou $et$
$al$. \cite{Shou-OL-2011}. They retained the terms of the Taylor
expansion of the light-induced refractive index up to the second
order whose coefficient is the on-axis light intensity. The phase
shift rate in (1+2)D LG was predicted to be much smaller
than the result based on SN model, but is still more than one order
larger than that in the local media. More meaningful, Shou $et$
$al$. observed a linear modulation of the soliton power on the phase
shift of the SNSOS \cite{Shou-OL-2011}, which coincides with Guo's
prediction, indicating that the nonlocality enhancement to the phase
shift of SNSOS stems from the fact that the light-induced refractive
index, which directly contributes to the phase shift, is induced not
by the light intensity but by the power of the whole beam.

In this paper, we investigate the phase shift of SNSOS in (1+1)D
LG in the formalism of perturbation theory. The perturbation
solution of the fundamental soliton is obtained under the second
approximation. The result indicates that the phase shift of SNSOS in
(1+1)D LG is proportional to the degree of nonlocality which
is at least one order larger than the result for the local solitons. It
will also be shown how the degree of nonlocality affects, or explicitly speaking, enhances the phase
shift rate in different nonlocal systems.

\section{The fundamental strongly nonlocal soliton solution under the second approximation}

We consider a (1+1)D LG with thermal nonlocal nonlinear
response occupying the region $-L\leq x\leq L$. The propagation
behavior of a light beam $\emph{u}$ propagating along the $z$ axis
is governed by the NNLSE, coupled to the Poisson equation describing
the light-induced nonlinear refractive index variation \emph{N},
\begin{equation}\label{qblNNLSE-1}
i\frac{\partial u}{\partial
z}+\frac{1}{2}\frac{\partial^{2}u}{\partial x^{2}}+Nu=0,
\end{equation}
\begin{equation}\label{qblzsl-1}
\frac{d^{2}N}{dx^{2}}=-|u|^{2}.
\end{equation}
The nonlocal response function in (1+1)D LG under the
first-kind boundary condition $N(\pm L)=0$ can be given as
\cite{Arfken-1985}
\begin{equation}\label{xiangying-1}
 G(x,\xi)=\left\{
            \begin{array}{ll}
              \frac{(x+L)(\xi-L)}{2L},\qquad (x\leq \xi) \\\\
              \frac{(\xi+L)(x-L)}{2L}.\qquad (\xi\leq x)
            \end{array}
          \right.
\end{equation}
According to the Green function method, the nonlinear refractive
index in LG can be written in the form of
\begin{equation}
N(x)=-\int_{-L}^{L}G(x,\xi)|u(\xi,z)|^{2}d\xi.
\end{equation}

It is obvious that the response function in Eq.(\ref{xiangying-1})
is not differentiable at the source point $x=\xi$ (irregular) and therefore
cannot be dealt with SN model \cite{Guo-PRE-2004}. We use the
perturbation method, previously extended to solve the NNLSE by
Ouyang $et$ $al$. \cite{Ouyang-PRE-2006, Ouyang-PRA-2007}, to
calculate the fundamental soliton solution of the NNLSE. For the
soliton state $u(x,z)$, we have $|u(-x,z)|^{2}=|u(x,z)|^{2}$ and
$u(x,z)=u(x,0)$. On the analogy of the potential in quantum
mechanics which determines the state of the particle movement, we
define the nonlinearity-induced trapping `potential', explicitly the
light-induced refractive index, which can determine the beam
propagation behavior,
\begin{equation}\label{v-1}
V(x)=\int_{-L}^{L}G(x,\xi)|u(\xi,z)|^{2}d\xi.
\end{equation}
Then Eq.(\ref{qblNNLSE-1}) can be reduced to
\begin{equation}\label{qblNNLSE-2}
i\frac{\partial u}{\partial
z}+\frac{1}{2}\frac{\partial^{2}u}{\partial x^{2}}-V(x)u=0.
\end{equation}
Taking the Taylor's expansion of $V(x)$ at $x=0$, we obtain
\begin{equation}\label{vv}
V(x)=V_{0}+\frac{1}{2\mu^{4}}x^{2}+\alpha x^{4}+\beta x^{6}+\cdots,
\end{equation}
where
\begin{subequations}\label{vv-1}
\begin{equation}\label{v0-1}
V_{0}=V(0),
\end{equation}
\begin{equation}\label{mu-1}
\frac{1}{\mu^{4}}=V^{(2)}(0),
\end{equation}
\begin{equation}\label{alpha-1}
\alpha=\frac{1}{4!}V^{(4)}(0),
\end{equation}
\begin{equation}\label{beta-1}
\beta=\frac{1}{6!}V^{(6)}(0).
\end{equation}
\end{subequations}

In the strongly nonlocal case, $V(x)$ is effective mainly within the
beam region. Consequently the terms $\alpha x^{4}$ and $\beta x^{6}$
are, respectively, one and two orders of magnitude smaller than the
term $x^{2}/(2\mu^{4})$ \cite{Ouyang-PRE-2006} and then can be
viewed as the perturbations. By substituting Eq.(\ref{vv}) into
Eq.(\ref{qblNNLSE-2}) and neglecting the higher-order terms, we
obtain
\begin{equation}\label{qbl-1}
i\frac{\partial u}{\partial
z}=\left[-\frac{1}{2}\frac{\partial^{2}}{\partial
x^{2}}+V_{0}+\frac{1}{2\mu^{4}}x^{2}+\alpha x^{4}+\beta
x^{6}\right]u.
\end{equation}
Taking a transformation
\begin{equation}\label{bhuan-1}
u(x,z)=\phi(x)\exp[-i(\varepsilon+V_{0})z],
\end{equation}
we arrive at
\begin{equation}\label{qbl-2}
\left[-\frac{1}{2}\frac{d^{2}}{dx^{2}}+\frac{1}{2\mu^{4}}x^{2}+\alpha
x^{4}+\beta x^{6}\right]\phi=\varepsilon\phi.
\end{equation}

If $\alpha=0$ and $\beta=0$, Eq.(\ref{qbl-2}) reduces to the
well-known stationary Schr\"{o}dinger equation for a harmonic
oscillator. Following the perturbation method we obtain the
fundamental soliton solution under the second approximation
\begin{eqnarray}\label{guzijie-1}
&& \phi_{0}(A,\alpha,\beta,x) \approx  A\left(\frac{1}{\pi \mu^{2}}\right)^{1/4}\exp\left(-\frac{x^{2}}{2\mu^{2}}\right)\nonumber\\
&&\times\bigg[1+\alpha\left(\frac{9\mu^{6}}{16}-\frac{3\mu^{4}}{4}x^{2}-\frac{\mu^{2}}{4}x^{4}\right)\nonumber\\
&&+\alpha^{2}\left(-\frac{1247\mu^{12}}{512}+\frac{141\mu^{10}}{64}x^{2}+\frac{53\mu^{8}}{64}x^{4}+\frac{13\mu^{6}}{48}x^{6}+\frac{\mu^{4}}{32}x^{8}\right)\nonumber\\
&&+\beta\left(\frac{55\mu^{8}}{32}-\frac{15\mu^{6}}{8}x^{2}-\frac{5\mu^{4}}{8}x^{4}-\frac{\mu^{2}}{6}x^{6}\right)\bigg],
\end{eqnarray}
and
\begin{equation}\label{canshu-1}
\varepsilon_{0}\approx\frac{1}{2\mu^{2}}+\frac{3\mu^{4}\alpha}{4}-\frac{21\mu^{10}\alpha^{2}}{8}+\frac{15\mu^{6}\beta}{8}.
\end{equation}

In the strongly nonlocal case, $\alpha$ and $\beta$ are very small,
and accordingly, so is the difference between the fundamental soliton
solution under the second approximation $\phi_{0}(A,\alpha,\beta,x)$
and that under the zeroth approximation $\phi_{0}(A,0,0,x)$. $V(x)$ can
be approximately given by
\begin{eqnarray}\label{VX}
V(x)&\approx& \int_{-L}^{L}G(x,\xi)\phi _{0}^{2}(A,0,0,\xi)d\xi\nonumber\\
&=&\frac{A^{2}}{2}\Bigg\{\frac{\mu}{\pi}\left[\exp\left(-\frac{x^{2}}{\mu^{2}}\right)-\exp\left(-\frac{L^{2}}{\mu^{2}}\right)\right]
\nonumber\\
&&-L\textup{erf}\left(\frac{L}{\mu}\right)
+x\textup{erf}\left(\frac{x}{\mu}\right)\Bigg\},
\end{eqnarray}
where
\begin{equation}\label{canshu-2}
\textup{erf}(x)=\frac{2}{\sqrt{\pi}}\int_{0}^{x}e^{-\xi^{2}}d\xi.
\end{equation}
Combining Eq.(\ref{vv-1}), we have
\begin{subequations}\label{vv-3}
\begin{equation}
A^{2}\approx\frac{\sqrt{\pi}}{\mu^{3}},
\end{equation}
\begin{equation}
V_{0}\approx\frac{\sqrt{\pi}}{2\mu^{3}}\Bigg\{\frac{\mu}{\sqrt{\pi}}\bigg[1-\exp\left(-\frac{L^{2}}{\mu^{2}}\right)\bigg]-L\textup{erf}\left(\frac{L}{\mu}\right)
\Bigg\},
\end{equation}
\begin{equation}
\alpha\approx -\frac{1}{12\mu^{6}},
\end{equation}
\begin{equation}
\beta\approx \frac{1}{60\mu^{8}}.
\end{equation}
\end{subequations}
Inserting Eq.(\ref{guzijie-1}) into Eq.(\ref{bhuan-1}), we find the
fundamental soliton solution in (1+1)D LG,
\begin{eqnarray}\label{uu-2}
&& u(x,z)\approx A\left(\frac{1}{\pi \mu^{2}}\right)^{1/4}\exp\left(-\frac{x^{2}}{2\mu^{2}}\right)\exp(i\gamma z)\nonumber\\
&
&\times\bigg[0.9649+a\frac{x^2}{\mu^2}+b\frac{x^4}{\mu^4}+c\frac{x^6}{\mu^6}+0.0002\frac{x^8}{\mu^8}\bigg],
\end{eqnarray}
where
$A^{2}\approx\frac{\sqrt{\pi}}{\mu^{3}},a=0.0386,b=0.0162,c=-0.0009,d=0.0002$,
and the phase shift rate is of the form
\begin{equation}\label{xiangyi-1}
\gamma
=-V_{0}-\varepsilon_{0}\approx\frac{1}{2\mu^{2}}\Bigg[\frac{\sqrt{\pi}L}{\mu}\textup{erf}\left(\frac{L}{\mu}\right)+\exp\left(-\frac{L^{2}}{\mu^{2}}\right)-1.87\Bigg].
\end{equation}
It is important to notice that in Eq.(\ref{uu-2}), $\mu$, defined in
Eq. (\ref{mu-1}), is visualized as the beam width. The power of the
soliton is approximatively given by
\begin{equation}\label{gonglv-1}
P=\int^{+\infty}_{-\infty}|u(x,z)|^2dx \approx
A^{2}\approx\frac{\sqrt{\pi}}{\mu^{3}}.
\end{equation}

In the above equations, $L$ plays the part of the characteristic
length $w_m$ of the response function, since $w_m$ of the LG
modeled by Eq.(\ref{qblzsl-1}) is intrinsically infinite but cut off
by its boundary \cite{Shou-OL-2009}. Therefore the ratio $L/\mu$
represents the degree of nonlocality $\alpha$. In the strongly
nonlocal limit, $\alpha\gg1$, we have
\begin{equation}\label{xiangyi-2}
\gamma \approx\frac{\alpha \sqrt{\pi}}{2\mu^{2}}
\end{equation}
It can be seen that the phase shift rate of SNSOS is proportional to
the degree of nonlocality, which is at least one order larger than
that for local solitons.

\begin{figure}[htbp]
 \centerline{\includegraphics[height=1.5in,width=3.4in]{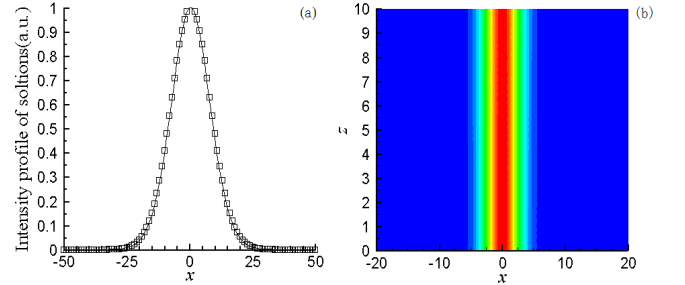}}\label{fig1}
 \caption{(a) The intensity profile of the (1+1)D soliton in LG for $\alpha = 150$.
 The squares represent the iterative solution and the solid line represents the perturbative solution expressed in Eq.(\ref{uu-2}).
 (b) Simulation of the soliton propagation where the perturbative solution (solid line in (a)) serves as the incident profile.}
\end{figure}

\begin{figure}[htbp]
\centerline{\includegraphics[width=.8\columnwidth]{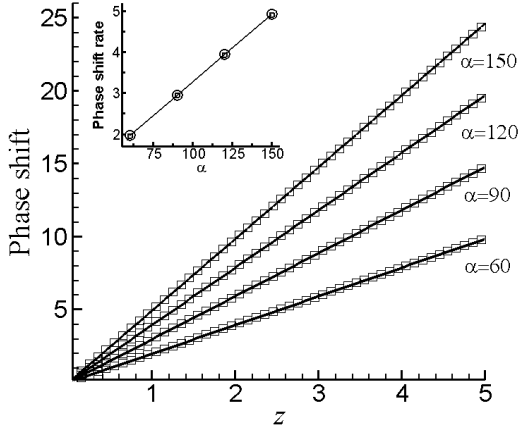}}\label{phase shift rate VS nonlocality}
\caption{Comparison of the phase shifts of SNSOS in (1+1)D LG with different degree of nonlocality
 $\alpha$ between the perturbative analytic results (solid lines) with the numerical results (squares).
 Inset shows the phase shift rate as a function of the degree of  nonlocality $\alpha$. Circles and squares are respectively calculated based on the analytical and numerical results. Solid line is provided as a guide to the eye.}
\end{figure}

Fig. 1(a) displays the intensity profile of the (1+1)D SNSOS in LG with $\alpha =150$. In the strongly nonlocal case, the
perturbative analytic result (solid line) is very close to the numerical result (squares) denoted
by the good agreement between them. Fig. 1(b) shows the
simulation of the light beam propagation in the form of soliton with an input amplitude profile
described by Eq.(\ref{uu-2}). The phase shift of the (1+1)D SNSOS in
LG versus the propagation distance is manifested in Fig. 2.
Under different conditions of nonlocalities, the higher the degree
of nonlocality, the faster the phase shift gets. The phase shift
rates of SNSOSs are obtained by calculating the slopes of the data
in Fig.2. It suggests that SNSOS experiences $\pi$ phase shift
within one Rayleigh distance in (1+1)D LG. The inset of Fig.
2 reveals the phase shift rate changes along with the degree of the
nonlocality. It visualizes that the enhancement-effect of the
nonlocality on the phase shift rate acts linearly and effectively.
One can, therefore, obtain faster phase shift
by directly enlarging the size of the LG since the
degree of nonlocality of LG is determined by the glass size.

\section{Discussion}

In the previous section, we investigate the phase shift of SNSOS in
(1+1)D LG and find that the phase shift rate is proportional
to the degree of nonlocality. A quantitative comparisons of phase
shift rates of spatial solitons in (1+1)D and (1+2)D materials with
different response functions are represented in Table 1
\cite{Guo-PRE-2004, Ouyang-PRE-2006, Ouyang-PRA-2007, Shou-OL-2011,
Aitchison-OL-1990}. Fig. 3 gives an illustration of the phase shift
rates of solitons versus the degree of nonlocality $\alpha$ in
different local and nonlocal systems. There are several features
that should be emphasized. First of all, solitons propagating in
material with nonlocal nonlinear response have much faster phase
shift rate. We call this `nonlocality-enhanced phase shift'. In
nonlocal media with Gaussian response functions,
nonlocality-enhancing factor is $\alpha^{2}$, which is more than 100,
in the strongly nonlocal cases \cite{Guo-PRE-2004}. In LG, the nonlocality-enhancing factors are much smaller but still
over 10 for the lower limit of the strong nonlocality
\cite{Shou-OL-2011}. Second, the nonlocality-enhancing factors
present different forms in (1+1)D and (1+2)D systems. Generally speaking, compared with higher dimensional SNSOSs, lower dimensional SNSOSs have much faster phase shift rates. Specifically, in (1+1)D NLC
and LG, phase shift rates have the same expressions which are
proportional to the degree of nonlocality $\alpha$
\cite{Ouyang-PRE-2006}. This indicates the phase shift rates of
SNSOS are more than one order of magnitude faster than those for the
local ones. While in the (1+2)D NLC and LG, the phase shift rate
takes the form of natural logarithm of $\alpha$ \cite{Shou-OL-2011,
Ouyang-PRA-2007}. Smaller although than that in the
lower-dimensional nonlocal media, nonlocality-enhancing factor is
still one order larger than the result for local solitons in LG, and
more that 5 times larger than the result for local solitons in NLC.
The nonlocality-enhancement-effect on the phase shift of SNSOS
originates from the fact that, the refractive index, directly
contributing to the phase shift, is induced not by the light
intensity but, thanks to the `averaging-effect' of the nonlocality,
by the power of the whole beam.

\renewcommand\arraystretch{1.5}

\begin{table}
\footnotesize \scriptsize \setcaptionwidth{3.5in}
\caption{\scriptsize{Phase shift rate of spatial solitons in (1+1)D
and (1+2)D materials with different response functions}} \centering
\begin{threeparttable}

\begin{tabular}{|c|c|p{66pt}|c|}
\hline \hline Dimension & Nonlocality  & Material & Phase shift
rate\\
\hline
\multirow{4}{*}{(1+1)D} & Local  &Local media   & $1/L_{R}$ \cite{Aitchison-OL-1990}\\
\cline{2-4}
& \multirow{3}{*}{Nonlocal} & \scriptsize{Media with Gaussian response}  & $\alpha^{2}/L_{R}$ \cite{Guo-PRE-2004}\\
\cline{3-4}
& & NLC  & $\sqrt{\pi}\alpha/L_{R}$ \cite{Ouyang-PRE-2006}\\
\cline{3-4}
& & LG  & $\sqrt{\pi}\alpha/L_{R}$ \\
\hline
\multirow{4}{*}{(1+2)D} & Local  &Local media  & unstable \\
\cline{2-4}
& \multirow{3}{*}{Nonlocal} & \scriptsize{Media with Gaussian response} & $\alpha^{2}/L_{R}$ \cite{Guo-PRE-2004}\\
\cline{3-4}
& & NLC  &$(8.6{\rm ln}\alpha-4.3)/\pi L_{R}$ \cite{Ouyang-PRA-2007}\\
\cline{3-4}
& & LG  & $(2{\rm ln}\alpha+6.24)/L_{R}$ \cite{Shou-OL-2011}\\
\hline \hline
\end{tabular}
\begin{tablenotes}
        \footnotesize \scriptsize
        \item[1]  $L_{R}$ is the Rayleigh distance.
        \item[2]  $\alpha$ is the degree of nonlocality defined as
        the ratio of the characteristic length of the response function to the beam
        width. In LG, the characteristic length of the response
        function is the medium size \cite{Shou-OL-2011}.
              \end{tablenotes}
\end{threeparttable}
\end{table}

\begin{figure}[htbp]
\centerline{\includegraphics[width=.8\columnwidth]{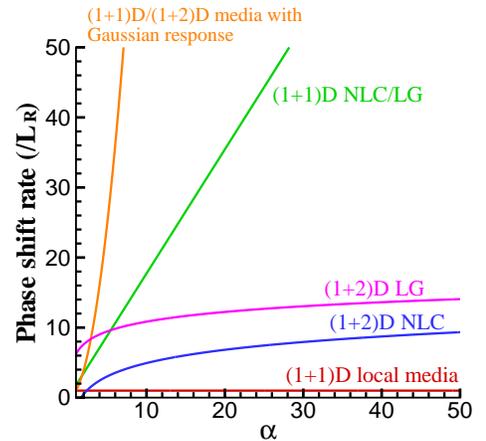}}
\caption{Comparison of the phase shift rates of solitons versus the degree of nonlocality $\alpha$ in different local and nonlocal systems. The solid curves are the phase shift rates, normalized by Rayleigh distance $L_{R}$ in, from top to bottom, (1+1)D or (1+2)D media with Gaussian response, (1+1)D NLC or LG, (1+2)D LG, (1+2)D NLC, (1+1)D local media, respectively.}
\end{figure}

\section{Conclusion}
Using the perturbation method, we investigate the large phase shift
of SNSOS in (1+1)D LG. The perturbative solution of the
fundamental soliton under the second approximation suggests that the
phase shift rate of (1+1)D SNSOS is proportional to the degree of
the nonlocality, which is at least one order faster than that of its local
counterpart. This facilitates a $\pi$-phase-shift within one Rayleigh distance in (1+1)D LG. The nonlocality-enhancement-effect on the phase shift of SNSOS is an important and intrinsic feature of nonlocality, which, although works differently in different nonlocal systems, leads to a much faster, generally one order faster, phase shift rate of SNSOS than that of the local counterpart. Phase shift is very important for modification, manipulation, and
control of optical field based on the principle of interference. The
nonlocality-enhancement to the phase shift might be of great
potential in applications based on the effective generation of large
phase shift.

\section*{Acknowledgment}

This research was supported by the National Natural Science
 Foundation of China (Grant No. 11274125), and the Natural Science Foundation of Guangdong Province of China (Grant No. S2012010009178).

\end{document}